\documentclass[10pt, conference, letterpaper]{IEEEtran}

\usepackage{graphicx}
\usepackage[misc,geometry]{ifsym} 
\usepackage{amsmath}
\usepackage{color}
\usepackage{hyperref} 
\usepackage[bottom]{footmisc}
\usepackage{caption}
\usepackage{supertabular}
\usepackage{afterpage}
\usepackage{url}
\usepackage{multicol}
\usepackage{multirow}
\usepackage{lscape}
\usepackage{float}
\usepackage{placeins}
\usepackage{afterpage}

\usepackage{afterpage}

\title{Interpretable by Design: MH-AutoML for Transparent and Efficient Android Malware Detection without Compromising Performance}

\author{
  \IEEEauthorblockN{
    \begin{minipage}[t]{0.32\linewidth}
      \centering
      Joner Assolin\textsuperscript{1}\\
      \footnotesize\textsuperscript{1}Federal University of Amazonas (UFAM)
    \end{minipage}
    \hfill
    \begin{minipage}[t]{0.32\linewidth}
      \centering
      Gabriel Canto\textsuperscript{1}\\
      \footnotesize\textsuperscript{1}Federal University of Amazonas (UFAM)
    \end{minipage}
    \hfill
    \begin{minipage}[t]{0.32\linewidth}
      \centering
      Diego Kreutz\textsuperscript{2}\\
      \footnotesize\textsuperscript{2}Federal University of Pampa (UNIPAMPA)
    \end{minipage}
  }
  
  \vspace{2mm} 
  
  \IEEEauthorblockN{
    \begin{minipage}[t]{0.32\linewidth}
      \centering
      Eduardo Feitosa\textsuperscript{1}\\
      \footnotesize\textsuperscript{1}Federal University of Amazonas (UFAM)
    \end{minipage}
    \hfill
    \begin{minipage}[t]{0.32\linewidth}
      \centering
      Hendrio Bragança\textsuperscript{1}\\
      \footnotesize\textsuperscript{1}Federal University of Amazonas (UFAM)
    \end{minipage}
    \hfill
    \begin{minipage}[t]{0.32\linewidth}
      \centering
      Angelo Nogueira\textsuperscript{2}\\
      \footnotesize\textsuperscript{2}Federal University of Pampa (UNIPAMPA)
    \end{minipage}
  }
  
  \vspace{2mm} 
  
  \IEEEauthorblockN{
    \centering
    \begin{minipage}[t]{0.32\linewidth}
      \centering
      Vanderson Rocha\textsuperscript{1}\\
      \footnotesize\textsuperscript{1}Federal University of Amazonas (UFAM)
    \end{minipage}
  }
}

\begin{document}

\maketitle

\begin{abstract}
Malware detection in Android systems requires both cybersecurity expertise and machine learning (ML) techniques. Automated Machine Learning (AutoML) has emerged as an approach to simplify ML development by reducing the need for specialized knowledge. However, current AutoML solutions typically operate as black-box systems with limited transparency, interpretability, and experiment traceability.
To address these limitations, we present MH-AutoML, a domain-specific framework for Android malware detection. MH-AutoML automates the entire ML pipeline, including data preprocessing, feature engineering, algorithm selection, and hyperparameter tuning. The framework incorporates capabilities for interpretability, debugging, and experiment tracking that are often missing in general-purpose solutions.
In this study, we compare MH-AutoML against seven established AutoML frameworks: Auto-Sklearn, AutoGluon, TPOT, HyperGBM, Auto-PyTorch, LightAutoML, and MLJAR. Results show that MH-AutoML achieves better recall rates while providing more transparency and control. The framework maintains computational efficiency comparable to other solutions, making it suitable for cybersecurity applications where both performance and explainability matter.
\end{abstract}


\section{Introduction}

The increasing sophistication of Android malware has led to a growing reliance on machine learning (ML) for detection, as traditional signature-based methods often struggle to keep pace with evolving threats. While ML offers promising capabilities, developing effective detection models requires substantial expertise in statistics, programming, and cybersecurity. This creates a significant barrier for practical adoption, particularly in environments where such specialized knowledge is scarce.

Automated machine learning (AutoML) \cite{truong2019automl} emerges as a potential solution by streamlining the ML workflow. By automating critical stages of the pipeline (i.e., data preprocessing, feature engineering, model selection, and hyperparameter optimization) AutoML reduces the technical burden, allowing practitioners to focus on domain-specific challenges rather than implementation details \cite{truong2019automl,wu2024data,zhang2025dream}. However, despite the rapid expansion of AutoML tools, their effectiveness and applicability in specialized domains like malware detection remain uncertain.

One key challenge lies in the diversity of AutoML solutions, which range from general-purpose frameworks to domain-specific implementations, making direct comparisons difficult \cite{zoller2021benchmark}. Additionally, critical aspects of the ML pipeline, such as explainable artificial intelligence (XAI), often receive insufficient attention \cite{carvalho2019machine,vilone2020explainable,barbudo2023eight,salehin2024automl}. Many existing AutoML tools also exhibit limitations in transparency, debugging support, transfer learning, scalability, and pipeline customization \cite{santu2021automl,barbudo2023eight,baratchi2024automated,tian2024automated,salehin2024automl}. These shortcomings are particularly problematic in security-sensitive applications like malware detection, where interpretability and operational trustworthiness are essential.

To bridge these gaps, we introduce 
MH-AutoML\footnote{This work significantly extends our previous contribution \cite{Assolin2024MHAutoML}, recognized as an outstanding artifact at SBSeg'24, by incorporating recent literature, evaluating performance on balanced datasets to address class imbalance, and introducing a novel interpretability assessment framework. Key enhancements include an expanded comparative analysis of AutoML tools using robust metrics (recall, MCC, execution time), a custom transparency evaluation model across five dimensions, and in-depth validation of \textit{MH-AutoML}'s effectiveness in imbalanced scenarios, demonstrating its consistent performance and superior explainability compared to existing solutions.}, 
a domain-specific AutoML framework tailored for Android malware detection. Unlike general-purpose solutions, MH-AutoML provides a fully automated pipeline that integrates data cleaning, feature engineering, model selection, and hyperparameter tuning, while also emphasizing transparency, interpretability, and debugging capabilities. These features, often overlooked in existing tools, are critical for ensuring that security analysts can validate and trust detection outcomes.

The main contributions of this work include:
\begin{enumerate}
    \item The design and implementation of MH-AutoML, an AutoML framework specifically optimized for Android malware detection. The framework supports the entire ML pipeline, including data preprocessing, feature engineering, algorithm selection, and hyperparameter tuning. MH-AutoML was developed with a focus on real-world applicability in cybersecurity, offering mechanisms that facilitate transparency and model validation;
    
    \item Novel support for transparency, interpretability, experiment tracking, and pipeline customization. To assess these aspects, we introduce a custom evaluation model based on scores derived from questions structured around five key dimensions: Functional Description, Statistical Analysis, Algorithmic Transparency, Interpretability, and Internal Analysis. This evaluation reveals MH-AutoML’s superiority over existing AutoML tools in terms of explainability and user insight;
    
    \item A comprehensive empirical evaluation comparing MH-AutoML against seven widely used general-purpose AutoML frameworks, including Auto-Sklearn\footnote{\url{https://automl.github.io/auto-sklearn}}, AutoGluon\footnote{\url{https://auto.gluon.ai}}, TPOT\footnote{\url{http://epistasislab.github.io/tpot}}, HyperGBM\footnote{\url{https://github.com/DataCanvasIO/HyperGBM}}, Auto-PyTorch\footnote{\url{https://automl.github.io/Auto-PyTorch}}, LightAutoML\footnote{\url{https://lightautoml.readthedocs.io}}, and MLJAR\footnote{\url{https://supervised.mljar.com}}. This evaluation uses both imbalanced and balanced datasets with unique samples to better assess the robustness of each tool under class imbalance conditions. We report comparative analyses on key metrics, such as recall, Matthews Correlation Coefficient (MCC), and execution time, across original, balanced, and imbalanced datasets. The results reinforce MH-AutoML's consistent performance and resilience in diverse and challenging data scenarios.
\end{enumerate}
Our results indicate that MH-AutoML delivers consistently high recall with competitive computational efficiency, making it a practical and explainable option for real-world cybersecurity applications.

\section{AutoML Pipeline}
\label{sec:pipeline}

\begin{figure}[!htb]
\centering
\includegraphics[width=0.48\textwidth]{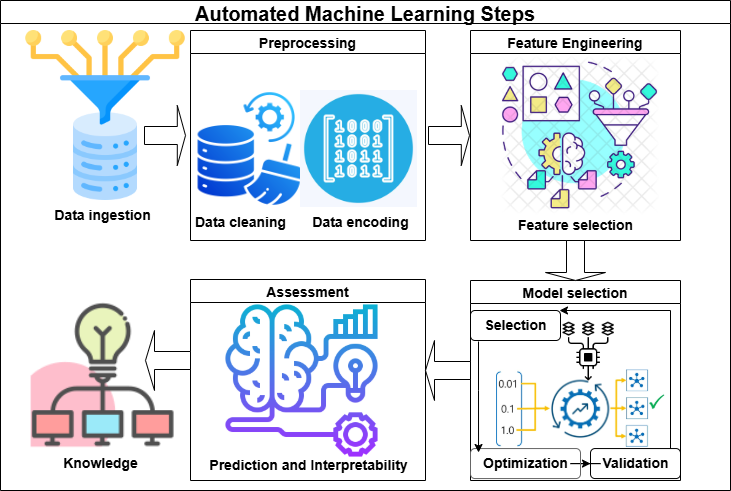}
\caption{AutoML execution steps.}
\label{fig_automl}
\end{figure}

The ML pipeline follows a structured sequence of steps that transform raw data into predictive models. Traditional approaches require manual execution of data preprocessing, feature selection, model choice, hyperparameter tuning, and evaluation, demanding significant technical expertise and iterative experimentation. AutoML automates these steps through algorithmic exploration of different technique combinations, as illustrated in Figure \ref{fig_automl} \cite{truong2019automl}.

\begin{figure}[!htb]
\centering
\includegraphics[width=0.45\textwidth]{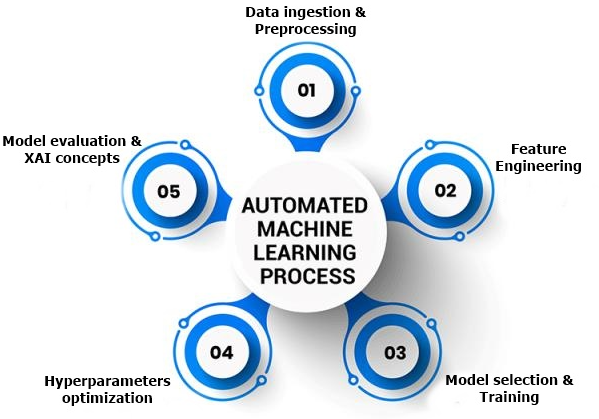}
\caption{AutoML process.}
\label{fig_automl_process}
\end{figure}

AutoML pipelines reduce human intervention by systematically searching for optimal configurations using methods like Bayesian optimization, genetic programming, and reinforcement learning \cite{hutter2019automated,doke2021survey,wu2024data,zhang2025dream}. This automation makes ML more accessible to non-experts while maintaining rigorous optimization standards. Figure \ref{fig_automl_process} details these automated steps and their implementation in typical AutoML systems.

Data preprocessing forms the foundation of the pipeline, where raw data undergoes transformations to improve quality and consistency. Common preprocessing tasks include:
\begin{itemize}
\item Imputation of missing values using statistical estimates;
\item Treatment of outliers through removal or dampening techniques;
\item Data transformations including scaling, normalization, and categorical encoding.
\end{itemize}
Automated systems select these preprocessing techniques through heuristic or optimization-driven approaches, reducing bias risks while maintaining model performance \cite{giovanelli2022data, bilal2022autoprep}.

Feature engineering builds upon preprocessed data by creating or modifying attributes to enhance predictive power. Typical operations include transforming existing features through encoding, discretization, or dimensionality reduction, and generating new features by combining attributes or extracting statistical properties.
AutoML implementations typically incorporate domain-specific knowledge, particularly in specialized applications like Android malware detection where features like permission patterns and obfuscation indicators prove valuable.

Model selection involves evaluating different algorithms against the problem requirements and dataset characteristics. AutoML systems automate this comparison across various model classes including decision trees, ensemble methods, neural networks, and linear models, using cross-validation to assess performance objectively.

The subsequent hyperparameter optimization phase systematically explores configuration spaces using methods like grid and random search for comprehensive exploration, and bayesian optimization and genetic algorithms for efficient search strategies.
These methods automatically identify parameter sets that maximize target metrics such as accuracy or F1-score \cite{wu2019hyperparameter,yang2020hyperparameter}.

Interpretability features allow understanding and validating model decisions through techniques like LIME \cite{ribeiro2016should} and SHAP \cite{lundberg2017unified}. In AutoML implementations, these explainability components help monitor pipeline construction and facilitate operational validation.

The final prediction analysis stage provides comprehensive model evaluation through:
\begin{itemize}
\item Performance metrics including accuracy, precision, recall, and AUC;
\item Visualizations like confusion matrices and ROC curves;
\item Comparative analysis of alternative approaches.
\end{itemize}
Automation in this phase enables continuous model improvement and maintains experiment histories as new data becomes available.

\section{Related Work}
\label{sec:related_work}

We organize our literature review to systematically explore the AutoML landscape through four interconnected sections. In Section \ref{sec_bench_studies}, we examine comparative studies that evaluate existing AutoML tools across various domains and performance metrics, establishing benchmarks for assessing tool capabilities. Building on this foundation, Section \ref{sec_proposed_automl} documents novel approaches that advance current AutoML methodologies, while also cataloging publicly available tools to provide practical references for practitioners and researchers.

Section \ref{sec_interp_analy} focuses on interpretability techniques and their implementation in AutoML systems, addressing the growing need for explainability. 
This organizational structure guides readers through a logical progression from performance evaluation to methodological innovations, followed by interpretability considerations and practical tool overviews.

Our approach helps readers first understand evaluation standards, then explore emerging developments, examine explainability solutions, and finally discover available implementations. Together, these sections provide a comprehensive perspective on the current AutoML ecosystem while identifying key research directions and practical applications.

\subsection{Benchmarking}
\label{sec_bench_studies}
 
We organize our analysis of AutoML benchmarking studies in Table \ref{tab_automl_benchmark}, which compares existing tools across various domains and performance metrics. These studies naturally divide into two groups based on their data sources: those using real-world datasets and those employing synthetic data.

\begin{table*}[!htp]
\caption{Benchmarking studies on AutoML tools and frameworks.}
\label{tab_automl_benchmark}
\scriptsize
\centering
\resizebox{0.99\textwidth}{!}{
\setlength{\tabcolsep}{4pt}
\renewcommand{\arraystretch}{1.3}
\begin{tabular*}{.948\textwidth}{@{}p{1.8 cm} p{0.8cm} p{1.5cm} p{0.8cm} p{0.5cm} p{5cm} p{2.7cm} p{1.5cm}@{}}
\hline \hline
\textbf{Ref.} & \textbf{Datasets} & \textbf{Domain} & \textbf{Type} & \textbf{Nº Tools} & \textbf{Compared Tools} & \textbf{Metrics} & \textbf{Validation}\\
\hline
\cite{truong2019towards} & 300 & Non-specific & Real data & 7 & Ludwig, H2O-Automl, TPOT, Darwin, Auto-sklearn, Auto-Keras, Auto-ml& Accuracy, MSE, F1 score & Holdout \\ \hline

\cite{kundu2021empirical} & 2 & Non-specific & Real data & 2 & 
AutoGluon-Tabular, Microsoft NNI & AUC, ROC curve & Holdout 
\\\hline

\cite{ferreira2021comparison} & 12 & Non-specific & Real data & 8 & Auto-Keras, Auto-PyTorch, Auto-Sklearn, AutoGluon, H2O AutoML, rminer AutoML, TPOT, TransmogrifA & MAE, AUC, F1 score & 10-fold Cross validation \\ \hline

\cite{gijsbers2024amlb} & 104 & Non-specific & Real data & 10 & AutoGluon, Auto-Sklearn 1/2, FLAML, GAMA, H2O AutoML, LightAutoML, MLJAR, NaiveAutoML, TPOT & AUC, log loss, RMSE, time & Cross validation \\  \hline

\cite{naser2023machine} & 6 & Non-specific & Real data & 5 & BigML, DataRobot, Dataiku, Exploratory, RapidMiner & Accuracy, AUC, Logloss, ROC, RMSE, MAE, R2 & 10-fold Cross validation \\ \hline

\cite{wever2021automl} & 24 & Non-specific & Real data & 5 & SMAC, Hyperband, BOHB, GGP, HTN-BF & Execution time, F1 score & 10-fold Cross validation \\ \hline

\cite{oliveira2024benchmarking} & 3 & Image processing & Real data & 3 & Google Vertex AI, NVIDIA TAO, AutoGluon & AP, mAP & Cross validation \\ \hline

\cite{hariri2024benchmarking} & 9 & Civil Engineering & Real data & 1 & -- & MAE, MSE, RMSE, R2, RMSLE, MAPE & 10-fold Cross validation \\ \hline

\cite{sirt2025comprehensive} & 3 & Non-specific & Synthetic data & 3 & Manual FE, Standard AutoML, Transfer Learning AutoML & Accuracy, F1-score, Sensitivity, Specificity, Time, Memory & Cross validation \\ \hline

\cite{da2024benchmarking} & 100 & Non-specific & Synthetic data & 4 & AutoML4Clust, Autocluster, cSmartML, ML2DAC & Adjusted rand index, Silhouette, Time, Memory & Cross validation \\ \hline

\cite{trirat2024automl} & 14 & Non-specific & Synthetic data & 5 & Human Models, AutoGluon, GPT-3.5, GPT-4, DS-Agent & Composite score, Norm. perf. & Cross validation \\ 
\hline \hline
\end{tabular*}}
\end{table*}

The real-world data studies span multiple application domains, ranging from general machine learning tasks to specialized areas such as civil engineering~\cite{oliveira2024benchmarking} and image processing~\cite{hariri2024benchmarking}. These investigations show substantial variation in their experimental design, particularly in the number of datasets used. Some studies employ extensive collections, with \cite{truong2019towards} evaluating 300 datasets, \cite{gijsbers2024amlb} analyzing 104, and \cite{ferreira2021comparison} working with 12. These comprehensive evaluations typically compare 7 to 10 different tools using standard performance metrics like AUC, F1-score, and RMSE. Other researchers like \cite{kundu2021empirical}, \cite{naser2023machine}, and \cite{wever2021automl} take more focused approaches with fewer datasets but similar evaluation frameworks. In domain-specific applications, we observe the use of specialized metrics such as mAP, R², and the Silhouette Coefficient to better assess performance in particular contexts.

Studies using synthetic data \cite{sirt2025comprehensive, trirat2024automl,da2024benchmarking} tend to work with smaller datasets (typically 3 to 100 samples) but explore more specialized aspects of AutoML systems. These investigations examine capabilities like feature engineering, transfer learning, and clustering performance, while expanding their evaluation criteria to include computational efficiency measures such as runtime and memory usage. Tools like AutoML4Clust and GPT-based models feature prominently in these synthetic data studies. Across both categories, we observe the frequent use of platforms such as AutoGluon and Auto-Sklearn, along with a clear progression in validation methods from basic holdout approaches to more sophisticated cross-validation techniques.

Our analysis reveals distinct benchmarking approaches tailored to different research goals. Real-world studies emphasize practical applicability across diverse domains, while synthetic data investigations enable deeper analysis of specific system capabilities and computational characteristics. The recurring presence of certain tools across multiple studies indicates their established position in the AutoML landscape, and the evolving validation methodologies demonstrate increasing rigor in evaluation practices. These patterns provide valuable insights into current trends and future directions for AutoML benchmarking research.

\subsection{AutoML tools} 
\label{sec_proposed_automl}

In Table~\ref{tab_automl-proposals}, we systematically organize AutoML tools, distinguishing between those that include comparative evaluations and those introducing new solutions without benchmarking against existing approaches. This categorization helps reveal methodological differences between general-purpose and domain-specific AutoML solutions.

\begin{table*}[!htp]
\scriptsize
\caption{Survey of AutoML implementations: comparing tools and tasks, interpretability techniques, metrics, and validation methods across application domains}
\label{tab_automl-proposals}
\renewcommand{\arraystretch}{1.3}
\begin{tabular*}{\textwidth}{p{1.3cm}p{1.2cm}p{1.8cm}p{1.3cm}p{1.5cm}p{.6cm}p{.7cm}p{.5cm}p{1.5cm}p{1.5cm}p{1.5cm}}
\hline \hline
\textbf{Ref.} & \textbf{Application \& Datasets} & \textbf{Compared Tools} & \textbf{Tasks} & \textbf{Interpretability Method} & \textbf{Trace.} & \textbf{Transp.} & \textbf{Debug} & \textbf{Metrics} & \textbf{M. Sys} & \textbf{Validation Method} \\\hline

\cite{zimmer2000auto} & General (35) & 
Auto-Keras, Auto-Sklearn, Hyperopt-sklearn, Auto-Net2.0, AutoGluon & 
Classification, Regression, Forecasting & 
-- & 
-- & T1, T2 & D1,D2 & 
Avg. validation accuracy, Mean relative regret & 
No & Holdout \\
\hline

\cite{erickson2020autogluon} & General (50) & AutoGluon & 
Classification, Regression & 
Feature importance & 
-- & T1, T2 & D1,D2 & 
RMSLE, R², MAE, Log-loss, AUC, Gini & 
Execution Time, Memory Utilization & 10-fold CV \\
\hline

\cite{ledell2020h2o} & General (44) & 
XGBoost GBM, H2O RF, H2O XRT, H2O GBM, H2O DL, H2O GLM & 
Classification, Regression & 
Feature importance, PD, TreeSHAP, ICE & 
H2O Flow, MLflow & T1, T2 & D1,D2 & 
AUC, Logloss, MPCE, RMSE & 
Execution Time, Memory Utilization, Disk Usage, Network Usage & 10-fold CV \\
\hline

\cite{molino2019ludwig} & General (--) & 
TensorFlow, Keras, PyTorch, Ludwig & 
Classification, Regression & 
-- & 
MLflow, Comet, WandB & T1, T2 & D1,D2 & 
-- & 
Execution Time, Memory Utilization, Disk Usage, Network Usage & -- \\
\hline

\cite{neto2020nasirt} & Infrared (3) & Auto-Keras & 
Classification, Regression, Clustering & 
Feature importance, IRT & 
-- & T1, T2 & D1 & 
Accuracy & 
No & Holdout \\
\hline

\cite{nasimian2024alphaml} & Medical (6) & -- & 
Classification, Regression, Clustering, Forecasting & 
Feature importance, Permutation importance, SHAP, LIME & 
-- & T1, T2 & D1 & 
HL, KMCE, CS, Acc, AUC, CK, F1 score, Jacc, MCC, NLR, NPV, Prec, Sens, Spec & 
No & 5-fold CV \\
\hline

\cite{ye2025integrated} & Geological (1) & -- & 
& SHAP, Feature importance & 
-- & -- & -- & 
EUR, CV score & 
No & 5-fold CV \\
\hline

\cite{kovalevsky2024automl} & Fitness (1) & -- & 
Classification & 
-- & 
-- & -- & -- & 
Accuracy, Precision, Recall, AUC, F-score & 
No & Holdout \\
\hline

\cite{singh2024automl} & Groundwater (7) & -- & 
Classification & 
-- & 
-- & -- & -- & 
R, RMSE, BIAS & 
No & CV \\
\hline

\textbf{This work} & Android Malware (9) & 
MlJar, AutoSklearn, AutoGluon, TPOT, LightAutoML, AutoPyTorch, HyperGBM & 
Classification & 
SHAP, LIME, Feature importance & 
MLflow & T1, T2 & D1,D2 & 
Recall, Accuracy, F1 Score, Precision, MCC, ROC AUC & 
Execution Time, Memory Utilization, Disk Usage, Network Usage & 
Holdout + 5-fold CV \\
\hline \hline
\end{tabular*}%
\end{table*}

Our analysis shows that general-purpose tools typically employ comparative evaluations against established frameworks. Works like \cite{zimmer2000auto,ledell2020h2o,erickson2020autogluon} use standardized validation methods including 10-fold cross-validation, with \cite{molino2019ludwig} as the notable exception in its evaluation methodology. While these tools demonstrate robust performance benchmarking, most provide limited interpretability features. Only \cite{ledell2020h2o} and \cite{erickson2020autogluon} incorporate explanation methods, with \cite{ledell2020h2o} uniquely combining TreeSHAP and Individual Conditional Expectation plots with MLflow-based version control to enhance transparency and reproducibility.

Domain-specific solutions for applications in medicine \cite{nasimian2024alphaml}, geology \cite{ye2025integrated}, and spectroscopy \cite{neto2020nasirt} exhibit distinct characteristics. These tools generally focus less on comparative evaluations but emphasize domain-appropriate interpretability features. Similar to the general-purpose category, \cite{ledell2020h2o} remains the only solution implementing both traceability and version control capabilities in this group.

We observe an apparent trade-off in current AutoML development between rigorous performance benchmarking and comprehensive interpretability features. General-purpose tools tend to prioritize establishing performance standards through systematic comparisons, while domain-specific solutions concentrate more on explanation capabilities tailored to their application contexts.

Examining operational requirements across all evaluated tools, we identify several limitations in meeting basic transparency and debugging standards. Only five tools satisfy fundamental transparency criteria including documentation of algorithm relationships (T1) and hyperparameter configurations (T2). Fewer provide complete pipeline logs (D1) and specialized error reports (D2) for debugging purposes. The subset offering extended system monitoring (RAM, disk I/O, network metrics) shrinks to just three implementations: \cite{neto2020nasirt, erickson2020autogluon, ledell2020h2o}, indicating a significant gap in production-ready features.

These findings suggest that while the AutoML field is making progress in both performance benchmarking and domain-specific interpretability, comprehensive solutions addressing both aspects remain scarce. The limited availability of operational features like monitoring and debugging in most tools points to an important area for future development to enable effective real-world deployment.

In Table~\ref{tab_automl_tools}, we present an overview of publicly available AutoML tools, organized in six criteria. We identify each tool by name, specify its license type (open-source or proprietary), note the primary programming language used, indicate CLI availability, describe its application scope (general-purpose or domain-specific), and list the developing organization.

\begin{table*}
\centering
\scriptsize
\caption{Overview of publicly available AutoML tools: license, programming language, CLI, and scope.}
\renewcommand{\arraystretch}{1.5}
\begin{tabular*}{\textwidth}{@{} p{2.5cm} p{2.5cm} p{2.5cm} p{2.5cm} p{3.5cm} p{2.5cm}@{}}
\hline \hline
\textbf{Tool} & \textbf{License} & \textbf{Language} & \textbf{Native CLI?} & \textbf{Scope} & \textbf{Company/Org} \\ \hline
Auto-Keras & MIT & Python & Yes & General purpose & DATA Lab \\ \hline
Auto-Sklearn & BSD-3 & Python & Yes & General purpose & Uni Freiburg \\ \hline
AutoGluon & Apache 2.0 & Python & Yes & General purpose & AWS \\ \hline
H2O AutoML & Apache 2.0 & Java & Yes & General purpose & \href{http://H2O.ai}{H2O.ai} \\ \hline
LightAutoML & Apache 2.0 & Python & Yes & General purpose & Sberbank \\ \hline
TPOT & LGPL & Python & Yes & General purpose & UPenn \\ \hline
Ludwig & Apache 2.0 & Python & Yes & General purpose & Uber \\ \hline
AutoML4Clust & MIT & Python & No & Clustering analyses & Academia \\ \hline
FLAML & MIT & .NET & No & General purpose & Microsoft \\ \hline
TransmogrifAI & BSD-3 & Scala & No & General purpose (structured data) & Salesforce \\ \hline
Darwin & Proprietary & Python & No & Computer Vision & V7 Labs \\ \hline
DataRobot & Proprietary & R & No & General purpose & DataRobot Inc. \\ \hline
MLJAR & Proprietary* & Python & Yes & General purpose & MLJAR \\ \hline
Vertex AI & Proprietary & Python & Yes & General purpose & Google \\ \hline
NVIDIA TAO & Proprietary & Python & Yes & Computer Vision & NVIDIA \\ \hline \hline
\end{tabular*}
\label{tab_automl_tools}
\end{table*}

Our analysis shows Python as the primary language for 12 of the 15 tools listed, aligning with its established position in data science and machine learning communities. Most tools adopt open-source models with modular architectures supporting customization and extension. Tools like Auto-sklearn, AutoGluon, and TPOT demonstrate this flexibility, allowing users to modify components from data preprocessing to model optimization.

We note that 10 of the 15 tools provide native CLI support, simplifying their integration into automated workflows. While most solutions target general-purpose applications, we identify specialized tools including Darwin and NVIDIA TAO for computer vision tasks, and AutoML4Clust for clustering analysis. The development ecosystem includes major technology companies (e.g., AWS, Google, Microsoft, NVIDIA), startups (e.g., V7 Labs), and academic institutions (e.g., Uni Freiburg, UPenn, DATA Lab).

This variety reflects both the maturity of the field and growing interest from various sectors. However, we note a gap: while most tools adopt general-purpose approaches, few specialize in particular domains. Darwin and NVIDIA TAO focus on computer vision, while AutoML4Clust addresses clustering tasks.

The predominance of generalist tools suggests opportunities for more specialized solutions. In this context, MH-AutoML emerges as a promising development for Android malware detection, incorporating domain-specific knowledge from cybersecurity and Android app analysis to improve feature selection, modeling, and result interpretation.

Again, we observe growing research on domain-specific AutoML tools \cite{oakes2024building,singh2024automl,krzywanski2024automl,schaller2025automl,zheng2023automl,barbudo2023eight,salehin2024automl,baratchi2024automated}. For example, AutoML-GWL \cite{singh2024automl} addresses groundwater level prediction challenges by automatically selecting and optimizing models while incorporating hydrogeological factors. The framework maintains interpretability through SHAP analysis and demonstrates strong performance across diverse aquifer systems, proving valuable for water resource management and drought prediction applications.

\subsection{Interpretability in AutoML}
\label{sec_interp_analy}

\begin{figure*}[!bt]
\captionsetup{type=table}
\centering
\caption{Overview of interpretability methods in AutoML-related studies.}
\label{tab_interpretability-methods}
\includegraphics[width=0.99\textwidth]{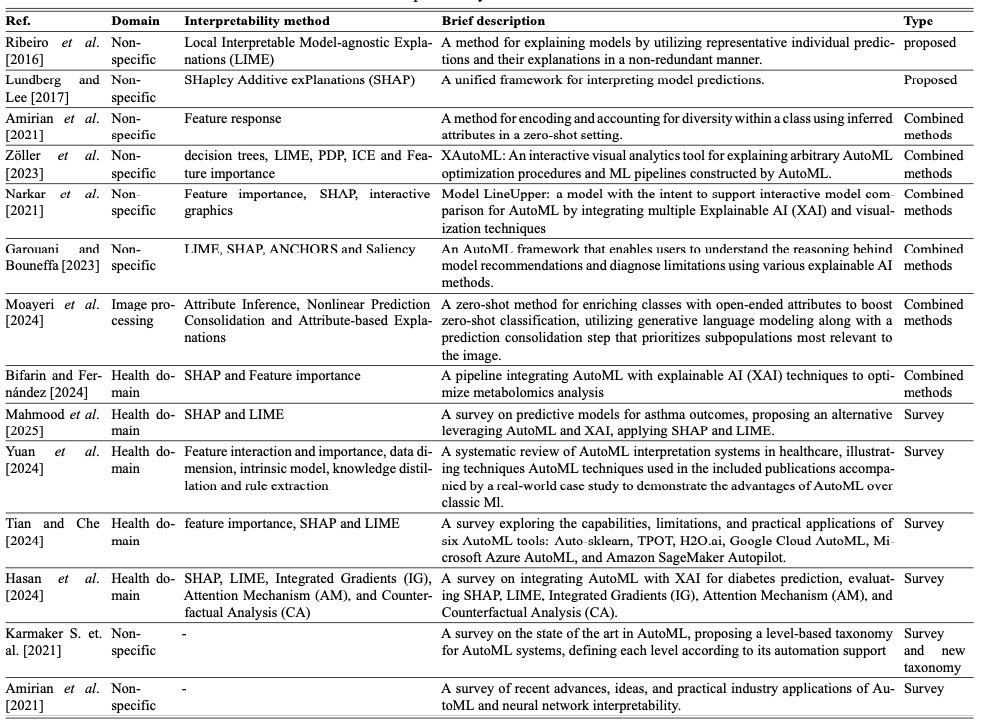}
\end{figure*}

We present in Table \ref{tab_interpretability-methods} a comprehensive overview of interpretability methods and tools, along with relevant survey literature. Our analysis reveals three main categories of contributions in this field. Foundational works like LIME and SHAP established core theoretical frameworks for model explanations. Subsequent studies such as \cite{zoller2023xautoml, bifarin2024automated, garouani2023unlocking} developed integrated approaches combining multiple techniques including SHAP, LIME, partial dependence plots, and feature importance. Comprehensive surveys including \cite{yuan2024automated, tian2024automated, karmaker2021automl} provide systematic reviews of current advancements while identifying research gaps.


  


General-purpose interpretability tools have evolved significantly since the introduction of LIME \cite{ribeiro2016should} and SHAP \cite{lundberg2017unified}. Recent work in this area focuses on combining existing techniques into novel methodologies \cite{amirian2021two}, developing visualization tools \cite{zoller2023xautoml, narkar2021model}, and creating unified frameworks \cite{garouani2023unlocking}. Domain-specific applications show particular innovation, with tailored solutions emerging for fields like image processing \cite{moayeri2024embracing} and healthcare \cite{bifarin2024automated}.

The survey literature reveals strong interest in healthcare applications, with multiple studies examining interpretability techniques in this domain \cite{hasan2024towards, mahmood2025interpretable, tian2024automated, yuan2024automated}. Broader surveys like \cite{karmaker2021automl} systematically review AutoML progress, while \cite{amirian2021two} synthesize recent innovations across applications. Our technical assessment shows that 62\% of studies use basic pre-model interpretability methods compared to 38\% employing advanced post-hoc approaches. Interactive explanation features appear in only 23\% of solutions despite their demonstrated value for model debugging.

This study extends previous work through systematic evaluation of seven major AutoML tools (MLJar, AutoSklearn, AutoGluon, TPOT, LightAutoML, AutoPyTorch, and HyperGBM) across nine Android-specific datasets. Our research represents the most comprehensive domain-specific comparison to date in terms of both tool coverage and dataset specialization. Additionally, we implement a unique combination of interpretability methods with transparency mechanisms, debugging support, performance tracking, and system monitoring metrics - features rarely found together in existing solutions but critical for production environments.

\section{MH-AutoML}
\label{sec:mh-automl}

In this work, we present MH-AutoML, a domain-specific tool for Android malware detection that addresses transparency, interpretability, and lifecycle control limitations in existing AutoML solutions. Our tool implements a complete AutoML pipeline that handles data cleaning, feature engineering, algorithm selection, hyperparameter tuning, and optimization.

MH-AutoML advances beyond conventional AutoML by integrating experiment tracking, debugging, and interpretability features throughout the pipeline, enabling users to inspect internal operations rather than just receiving final performance metrics.


\subsection{General Architecture and Tool Design}

\begin{figure*}[!bt]
\centering
\includegraphics[width=0.90\textwidth]{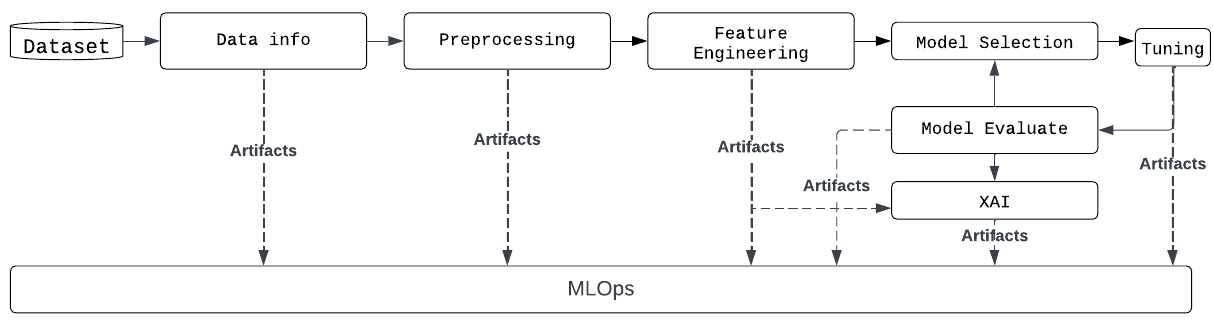}
\caption{Pipeline of MH-AutoML.}
\label{fig_fluxo}
\end{figure*}

Figure~\ref{fig_fluxo} illustrates MH-AutoML's workflow, showing the key stages of the ML process, including data preprocessing, feature engineering, and model selection with tuning.

The initial Data Info stage performs dual exploratory analysis, examining both the computational environment (OS, hardware resources, and network metrics) and dataset characteristics (dimensionality, data types, class balance, and integrity measures).

During Preprocessing, we address data inconsistencies through normalization and transformation. 
The current version handles common issues like missing values, incorrect data types, redundant entries, and improperly filled fields through outlier removal, duplicate elimination, missing value treatment, and optional one-hot encoding. 
The tool generates visual artifacts such as data quality heatmaps (see examples available on GitHub \cite{assolin2024mhautomlGitHub}).


The Feature Engineering stage implements dimensionality reduction techniques including PCA (\textit{Principal Component Analysis}) for feature extraction, ANOVA (\textit{Analysis of Variance}) for statistical selection, and LASSO (\textit{Least Absolute Shrinkage and Selection Operator}) for regularization. While PCA and ANOVA are common AutoML choices, LASSO was selected after thorough evaluation of over 20 feature selection methods across more than 10 heterogeneous datasets, showing superior generalization and robustness \cite{sfs2024vanderson}.


For Model Selection, we consider the problem requirements and dataset characteristics when choosing algorithms. Our implementation uses \texttt{VotingClassifier} to combine predictions from \texttt{LightGBM}, \texttt{KNN}, \texttt{CatBoost}, \texttt{RandomForestClassifier}, \texttt{DecisionTreeClassifier}, and \texttt{ExtraTreesClassifier}. This ensemble approach improves robustness, and the architecture supports easy integration of additional algorithms like \texttt{SVM}.

MH-AutoML includes built-in interpretability methods across its model portfolio. While \texttt{DecisionTreeClassifier} offers inherent interpretability through its rule-based structure, ensemble methods like \texttt{CatBoost} and \texttt{LightGBM} provide feature importance analysis despite their greater complexity. \texttt{KNN} and \texttt{ExtraTreesClassifier} present intermediate interpretability levels, with the understanding that ensemble models may require more effort to fully interpret in feature-rich scenarios.

The Model Evaluation stage employs multiple metrics to assess performance:

\begin{itemize}
\item Precision: Correct positive predictions relative to total positive predictions
\item Recall: Correct positive predictions relative to actual positives
\item Accuracy: Correct predictions across all classes
\item F1-score: Harmonic mean of precision and recall
\item MCC (\textit{Matthews Correlation Coefficient}): Binary classification quality metric
\item Execution Time: Model training and evaluation duration
\item ROC AUC (\textit{Receiver Operating Characteristic Area Under Curve}): Class separation capability
\item Precision-Recall Curve: Performance trade-off visualization
\item CPU/RAM/Disk/Network Usage: Resource consumption metrics
\end{itemize}

During Tuning, we optimize hyperparameters for each selected algorithm using the Optuna library, with Recall prioritized for malware detection effectiveness. For model interpretation, we implement LIME \cite{ribeiro2016should} and SHAP \cite{lundberg2017unified} to analyze feature contributions.

MH-AutoML integrates with the open Machine Learning Operations (MLOps) platform\footnote{\url{https://ml-ops.org}} through MLflow\footnote{\url{https://github.com/mlflow/mlflow}}, enabling pipeline lifecycle management. This includes model versioning, artifact storage (files, images, code), and experiment comparison - features not typically found in other MLflow-integrated tools. Additional pipeline artifacts and examples are available on GitHub\footnote{\url{https://github.com/SBSegSF24/MH-AutoML}}.

\section{Methodology}
\label{sec_metodologia}

We organize our methodology into five main components to ensure a comprehensive evaluation of AutoML tools. First, we establish our tool selection criteria in Subsection \ref{subsec_selecao}, followed by the performance evaluation framework in Subsection \ref{subsec_avaliacao}. We then present our interpretability assessment system in Subsection \ref{subsec_interpretabilidade}, the experimental setup in Subsection \ref{subsec_config}, and finally the dataset characteristics in Subsection \ref{subsec_datasets}. This structured approach allows us to systematically evaluate both the technical capabilities and transparency aspects of each AutoML solution.

\subsection{Selection Criteria}
\label{subsec_selecao}

For our comparative analysis, we selected seven established AutoML tools that represent the current state of open-source AutoML solutions: Auto-Sklearn \cite{feurer2015efficient}, AutoGluon \cite{erickson2020autogluon}, TPOT \cite{le2020scaling}, HyperGBM \cite{hypergbm}, Auto-PyTorch \cite{zimmer2021auto}, LightAutoML \cite{vakhrushev2021lightautoml}, and MLJAR \cite{plonska2021mljar}. Our selection process followed four key criteria to ensure the relevance and quality of the tools included in our study:

\begin{itemize}
\item Complete implementation of all AutoML pipeline stages as defined by \cite{truong2019towards}
\item Open-source availability with permissive licenses
\item Active maintenance with recent updates and community support
\item Recognition in recent benchmarking studies \cite{ferreira2021comparison,karmaker2021automl,assolin2022droidautoml,bahri2022automl,weerts2023fairness}
\end{itemize}

\subsection{Performance Criteria}
\label{subsec_avaliacao}

Our performance evaluation framework focuses on three key aspects of AutoML tools: detection capability, classification quality, and computational efficiency. We measure these using Recall (true positive rate), MCC (Matthews Correlation Coefficient), and execution time respectively. Table \ref{tab_metricas} provides the complete specifications for all evaluation metrics used in our study.

We conduct our experiments using eight carefully selected datasets (Table \ref{tab_datasets}) that cover diverse feature types and problem characteristics. All datasets are publicly available on GitHub \cite{assolin2024mhautomlGitHub} along with their original sources. Our data preparation process includes several important steps: First, we partition each dataset into training (80

For model validation, we employ the Holdout method, which provides a straightforward yet effective way to evaluate model generalization while maintaining computational efficiency. This approach aligns with common practices in AutoML benchmarking and facilitates direct comparison across different tools.

\begin{table}[h]
\renewcommand{\arraystretch}{2.0}
\centering
\caption{Metrics for evaluating the performance of AutoML tools, where: TP = True Positives; TN = True Negatives; FP = False Positives; FN = False Negatives.}
\label{tab_metricas}
\begin{tabular}{lc}
\hline
\textbf{Metric} & \textbf{Formula}                                                      \\ \hline
Recall          & $\frac{TP}{TP+FN}$                                                    \\
MCC             & $\frac{TP\cdot TN - FP\cdot FN}{\sqrt{(TP+FP)(TP+FN)(TN+FP)(TN+FN)}}$ \\ \hline
\end{tabular}
\end{table}

\subsection{Interpretability and Transparency Criteria}
\label{subsec_interpretabilidade}

To systematically evaluate interpretability and transparency aspects, we developed a structured questionnaire based on established literature \cite{love2023explainable, arrieta2020explainable, mi2020review, carvalho2019machine, brannstrom2023transparency}. Our framework organizes questions into five key categories that address different dimensions of tool transparency and explainability.

The Functional Description category examines whether tools clearly document their pipeline stages, components, inputs, outputs, and execution processes. Statistical Analysis evaluates how tools present results and performance metrics, including class predictions, training history, feature importance, and visualization methods.

Algorithmic Transparency assesses the clarity of information provided about models and hyperparameters, including logging capabilities and data balancing approaches. The Interpretability category has three subcomponents: pre-model (feature processing explanations), in-model (intrinsic interpretability), and post-model (prediction analysis and model-agnostic methods).

Finally, Internal Analysis focuses on tools' abilities to visualize model structures and decision processes. Table \ref{tab_checklist} shows our complete evaluation framework with scoring details.

We developed a quantitative scoring model that assigns 0-2 points per question based on implementation quality (Not applicable=0, Partial=1, Total=2). We normalize category scores to a 0-100\% scale using the formula:
\[
S = \begin{cases} 
0 & \text{if } N = 0 \\
1 & \text{if } N = 1 \\
2 & \text{if } N \geq 2 
\end{cases}
\]
where $S$ is the normalized score, $P_i$ is the question score, $N$ is the question count, and $W$ is the maximum score per question (2). This approach enables fair comparison across tools while maintaining scoring consistency.

\begin{table*}[htbp]
\centering
\caption{Example of the Evaluation Model for Transparency and Interpretability of AutoML Tools.}
\label{tab_checklist}
\small
\centering
\renewcommand{\arraystretch}{1.3}
\begin{tabular*}{\textwidth}{@{}p{3cm} p{11.3cm}p{0.7cm}p{0.7cm}@{}}
\hline \hline
\textbf{Category} & \textbf{Question} & \textbf{Score} & \textbf{Score} \\
\hline
\multirow{4}{*}{Functional Description} 
 & Is it possible to identify the pipeline stages? &  2 & \\
 & Is it possible to identify the components used in each stage? &  2 & 100 \\
 & Are the inputs and outputs clear?  & 2 & \\
 & Is it clear how the process is executed in practice? & 2 & \\
\hline
\multirow{5}{*}{Statistical Analysis} 
 & Are the model's predictions presented for the classes (malware and benign)?  & 2 &  \\
 & Is the history of data used to train and test the model maintained?  & 2 & \\
 & In feature selection, is the user informed about the most relevant features?  & 2 & 100\\
 & Are the model's performance metrics presented?  & 2 & \\
 & Are there ways to visualize the results (graphs, tables)?  & 2 & \\
\hline
\multirow{4}{*}{Algorithmic Transparency} 
 & Is it possible to identify which models are used?  & 2 &  \\
 & Is it possible to identify warnings, information, and errors recorded in the system logs?  & 2 & 100\\
 & Is it possible to identify the hyperparameters of the generated models?  & 2 & \\
 & Is it possible to identify if the data is balanced?  & 2 & \\
\hline
\multirow{6}{*}{Interpretability} 
 & Is the reason for dimensionality reduction explained? (Pre-Model)  & 2 & 100 \\
 & Are there domain-specific dimensionality reduction techniques? (Pre-Model)  & 2 & \\
 & Are global interpretability methods available?  & 2 & \\
 & Does the tool offer models with intrinsic interpretability? (In-Model)  & 2 & \\
 & Is it possible to evaluate the difference between predictions and actual values?  & 2 & \\
 & Are there model-agnostic interpretability methods? (Post-Model)  & 2 & \\
\hline
\multirow{1}{*}{Internal Analysis} 
& Are there methods to visualize the model's structure? (Post-Model)  & 2 & 100 \\ 
\hline \hline
\end{tabular*}
\end{table*}

\subsection{Experiment Configuration}
\label{subsec_config}

We conducted all experiments on a dedicated server with an Intel Xeon E5649 processor (2x6 cores @ 2.53GHz), 94GB RAM, and Linux Mint 20.3. Our Python 3.10.12 environment included key libraries: NumPy 1.23.5, Pandas 1.5.3, and scikit-learn 1.1.1.

To ensure a fair comparison, we used each tool's default configuration, only adjusting execution time limits when necessary to complete model training. This approach maintains consistency while respecting each tool's recommended settings.

\subsection{Datasets}
\label{subsec_datasets}

Our evaluation uses nine Android malware datasets that represent different analysis approaches and feature types (Table \ref{tab_datasets}). The datasets vary in size from 141 to 24,883 features, covering permissions (P), API calls (A), intents (I), system commands (S), and other characteristics (O). Class distributions range from balanced sets (DefenseDroid with ~6,000 samples per class) to highly imbalanced collections (AndroCrawl with 86,574 benign vs 10,170 malicious samples).

\begin{table*}[!ht]
\begin{center}
\caption{Summary of Dataset Information.}
\label{tab_datasets}
\resizebox{\textwidth}{!}{
\renewcommand{\arraystretch}{1.3}
\begin{tabular}{c|c||c|c|c|c}
\hline \hline
\multirow{2}{*}{\textbf{Reference}} &
\multirow{2}{*}{\textbf{\textit{Dataset}}} &
\multicolumn{2}{c|}{\textbf{Features}} &
\multicolumn{2}{c}{\textbf{Original Distribution}} \\
\cline{3-6}
 & & \textbf{Qty.} & \textbf{Types} & \textbf{Ben.} & \textbf{Mal.} \\
\hline 
\cite{martin2016adroit} & Adroit & 166 & P & 8058 & 3418 \\
\hline
\cite{sisto2012androcrawl} & AndroCrawl & 141 & A(26), I(8), P(84), O(23) & 86574 & 10170 \\
\hline
\cite{mahindru2018android} & Android Permissions & 151 & P & 9077 & 17787 \\
\hline
\cite{colaco2021defensedroid} & DefenseDroid PRS & 2877 & P(1489), I(1388) & 5975 & 6000 \\
\hline
\cite{colaco2021defensedroid} & DefenseDroid APICalls & 4275 & A & 5222 & 5254 \\
\hline
\cite{yerima2018droidfusion} & Drebin-215 & 215 & A(73), P(113), S(6), I(23) & 9476 & 5555 \\
\hline
\cite{guerra2021kronodroid} & KronoDroid Emulator & 276 & P(145), A(123), O(8) & 35246 & 28745 \\
\hline
\cite{guerra2021kronodroid} & KronoDroid Real & 286 & P(146), A(100), O(40) & 36755 & 41383 \\
\hline
\cite{sbseg} & MH-100K Real & 24.883 & P(166), A(24.417), I(250) & 89.254 & 12.721 \\
\hline \hline
\end{tabular}
}
\caption*{\small [P] Permissions, [A] API Calls, [I] Intents, [S] System Commands, [O] Others.}
\end{center}
\end{table*}

These datasets come from established malware research including Adroit \cite{martin2016adroit}, AndroCrawl \cite{sisto2012androcrawl}, DefenseDroid \cite{colaco2021defensedroid}, Drebin-215 \cite{yerima2018droidfusion}, and KronoDroid \cite{guerra2021kronodroid}. The feature composition reflects different detection approaches, from permission-based analysis to comprehensive behavioral profiling. This diversity allows us to thoroughly evaluate AutoML tool performance across various data characteristics.

\section{Results}
\label{sec_resultados}

We organize our results into four key analytical dimensions:
(a) Evaluation of tool interpretability and transparency features;
(b) Performance assessment using MCC and Recall metrics on original datasets;
(c) Performance evaluation on balanced datasets with unique samples;
(d) Comparative analysis between results from original and balanced datasets.

This structure allows us to systematically examine both the explainability characteristics and predictive performance of each AutoML tool across different data conditions. The comparative dimension provides insights into how data balancing affects tool performance in malware detection tasks.

\subsection{Transparency and Interpretability Evaluation}

Our evaluation of AutoML tools reveals distinct patterns in their transparency and interpretability characteristics across five assessment categories. In Functional Description, all tools achieved perfect scores (100 points), demonstrating comprehensive documentation of their workflows and operations. The Statistical Analysis category shows more variation, with MH-AutoML and MLJar leading at 100 points, followed by AutoGluon and HyperGBM at 80 points. LightAutoML, Auto-Sklearn, TPOT, and AutoPyTorch scored between 60-70 points, primarily due to incomplete presentation of data history and feature importance.

Algorithmic Transparency presents a different landscape, where only MH-AutoML achieved full marks (100 points). AutoGluon, AutoPyTorch, LightAutoML, HyperGBM, and MLJar each scored 75 points, while Auto-Sklearn and TPOT lagged at 50 points. These lower scores primarily reflect insufficient documentation of class balancing techniques and system logs.

The Interpretability assessment shows MH-AutoML and MLJar as top performers with 58.33 points, followed by HyperGBM (50 points) and LightAutoML/AutoGluon (41.6 points). AutoPyTorch (33.33 points), Auto-Sklearn (25 points), and TPOT (8.33 points) demonstrate varying degrees of limitations in their interpretability features. These limitations include missing support for intrinsic interpretability methods, lack of domain-specific explanations, and absence of model-agnostic techniques like SHAP and LIME.

Internal Analysis results present TPOT, HyperGBM, and MLJar as the strongest performers with 100 points each, excelling in model structure visualization. AutoGluon, MH-AutoML, and Auto-Sklearn show moderate capabilities (50 points), while AutoPyTorch and LightAutoML require significant improvement in this aspect (0 points).

\begin{figure*}[!htb]
\centering
\includegraphics[width=.8\textwidth]{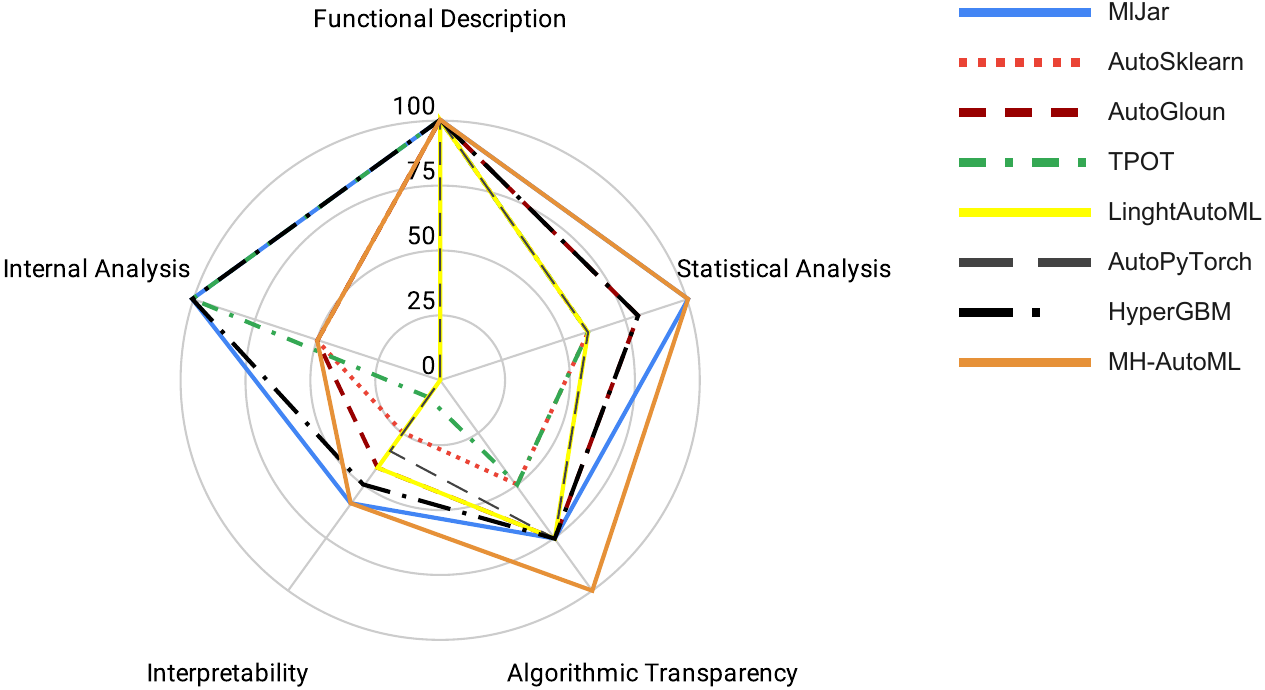}
\vspace{0.1cm}
\caption{Comparative evaluation of AutoML tools across five interpretability and transparency categories.}
\label{fig:AutoML_Feature_Score}
\end{figure*}

Figure \ref{fig:AutoML_Feature_Score} visually summarizes these findings, highlighting that while most tools perform well in basic functional documentation, many still need substantial improvements in model interpretability and internal structure visualization. These capabilities become particularly important in security-sensitive applications like Android malware detection, where understanding model decisions is as crucial as the predictions themselves. The results suggest that current AutoML tools vary significantly in their ability to provide transparent and interpretable solutions, with some tools demonstrating clear advantages in specific aspects of explainability.

\subsection{Tools Comparison on the Original Datasets}
\label{sec_resultados_datasets_originais}

\begin{figure*}[!htb]
\centering
\includegraphics[width=.75\textwidth]{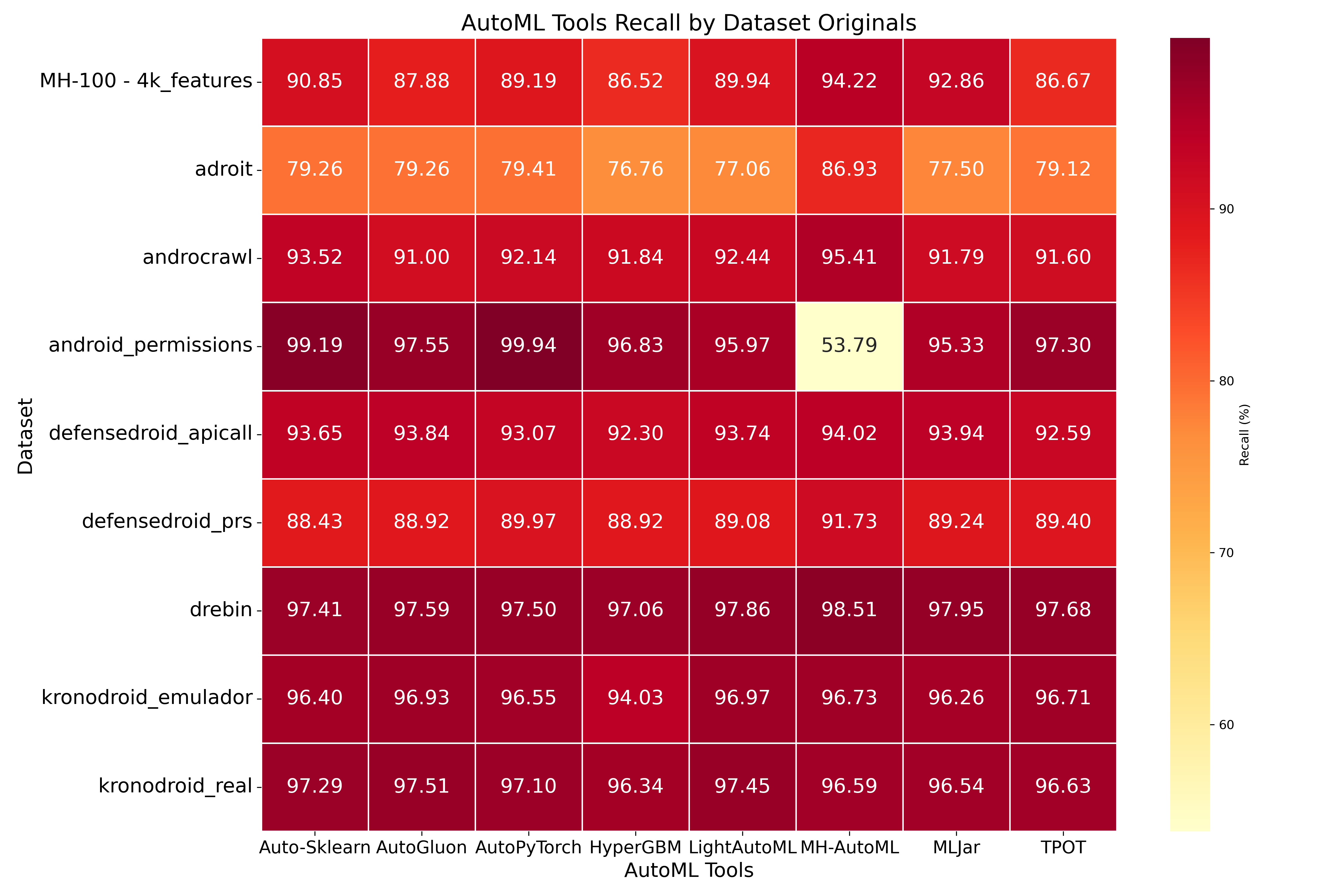}
\vspace{0.1cm}
\caption{Recall performance across datasets for all evaluated AutoML tools.}
\label{fig_desempenho_recall}
\end{figure*}

\begin{figure*}[!htb]
\centering
\includegraphics[width=.75\textwidth]{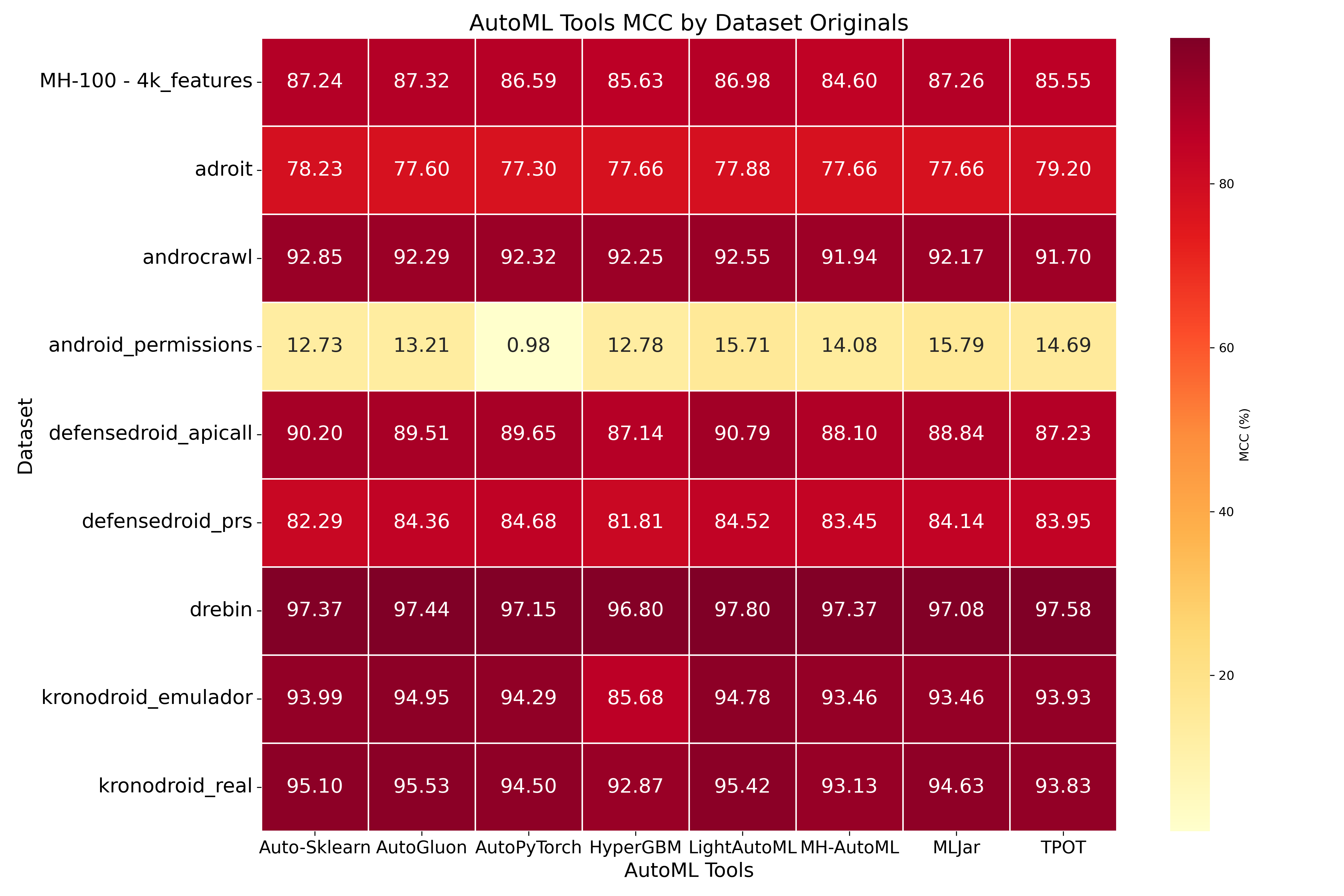}
\vspace{0.1cm}
\caption{MCC performance across datasets for all evaluated AutoML tools.}
\label{fig_desempenho_mcc}
\end{figure*}

\begin{figure*}[!htb]
\centering
\includegraphics[width=0.75\textwidth]{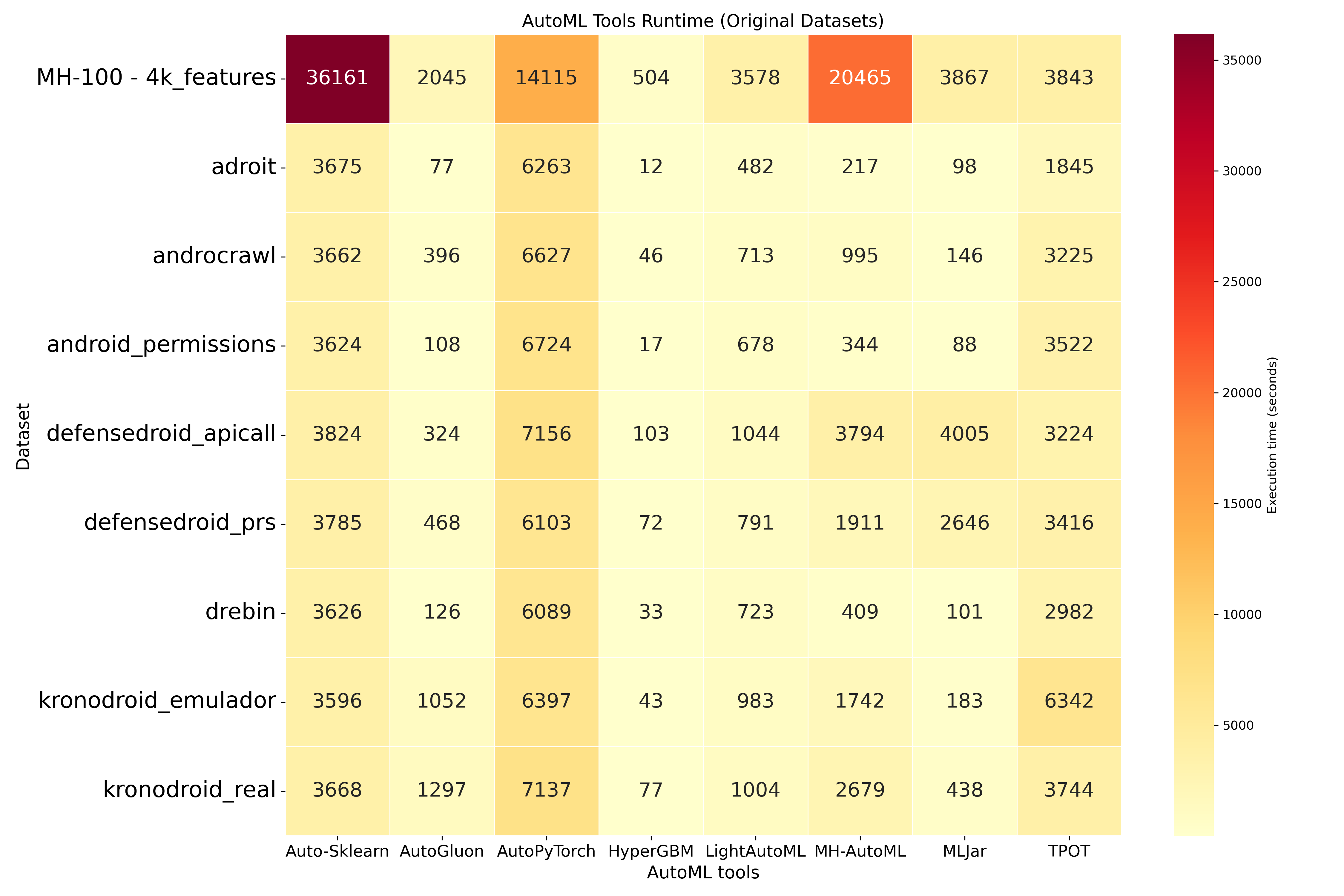}
\vspace{0.1cm}
\caption{Execution times across datasets for all evaluated AutoML tools.}
\label{fig_tempo_execucao}
\end{figure*}

We evaluate the performance of the 8 AutoML tools across 9 datasets using three key metrics: Recall, MCC, and execution time. Figures \ref{fig_desempenho_recall}, \ref{fig_desempenho_mcc}, and \ref{fig_tempo_execucao} present these results through comprehensive heatmaps that allow for visual comparison of tool performance.

Our analysis reveals three distinct performance groups among the tools. MH-AutoML, AutoGluon, and LightAutoML consistently achieve the best results across both evaluation metrics while maintaining reasonable execution times. In the DREBIN dataset, these tools demonstrate particularly strong performance, with LightAutoML reaching 97.80\% MCC and MH-AutoML achieving 98.51\% Recall, showcasing their ability to handle complex and imbalanced data effectively.

HyperGBM presents an interesting case of trade-offs between speed and accuracy. While it completes tasks quickly (e.g., 12 seconds for the Adroit dataset), its predictive performance often lags behind other tools. In the KronoDroid Emulator dataset, its 85.68\% MCC compares unfavorably to AutoGluon's 94.95\%, suggesting limitations in handling certain data characteristics despite its computational efficiency.

Execution time analysis reveals significant variation among tools. AutoGluon and MLJar maintain consistently low times, while AutoPyTorch, Auto-Sklearn, and TPOT require substantially longer processing durations without delivering proportional improvements in accuracy. For instance, AutoPyTorch needs 1 hour 46 minutes for the KronoDroid Emulator dataset, compared to AutoGluon's more efficient processing.

We observe some anomalous behaviors worth noting. MH-AutoML shows unexpectedly long execution times (5 hours 41 minutes) on the MH-100-4k dataset despite performing well on other datasets. Similarly, HyperGBM demonstrates inconsistent performance, achieving high Recall (96.83\%) but low MCC (12.78\%) on the Android Permissions dataset, indicating potential issues with false positives.

MH-AutoML emerges as particularly strong in Recall, leading in six of the nine datasets. However, its variable execution times suggest sensitivity to dataset complexity. LightAutoML shows the most consistent MCC performance, while AutoGluon provides the best balance between speed and accuracy across multiple datasets.

These results suggest that tool selection should consider both performance requirements and computational constraints. While some tools excel in specific metrics, the choice ultimately depends on whether the priority lies with predictive accuracy (Recall/MCC) or processing speed, particularly when dealing with large or complex malware detection datasets.

\subsection{Performance on Balanced Unique Sample Datasets}
\label{sec_resultados_datasets_unicas}

\begin{figure*}[!htb]
\centering
\includegraphics[width=0.75\textwidth]{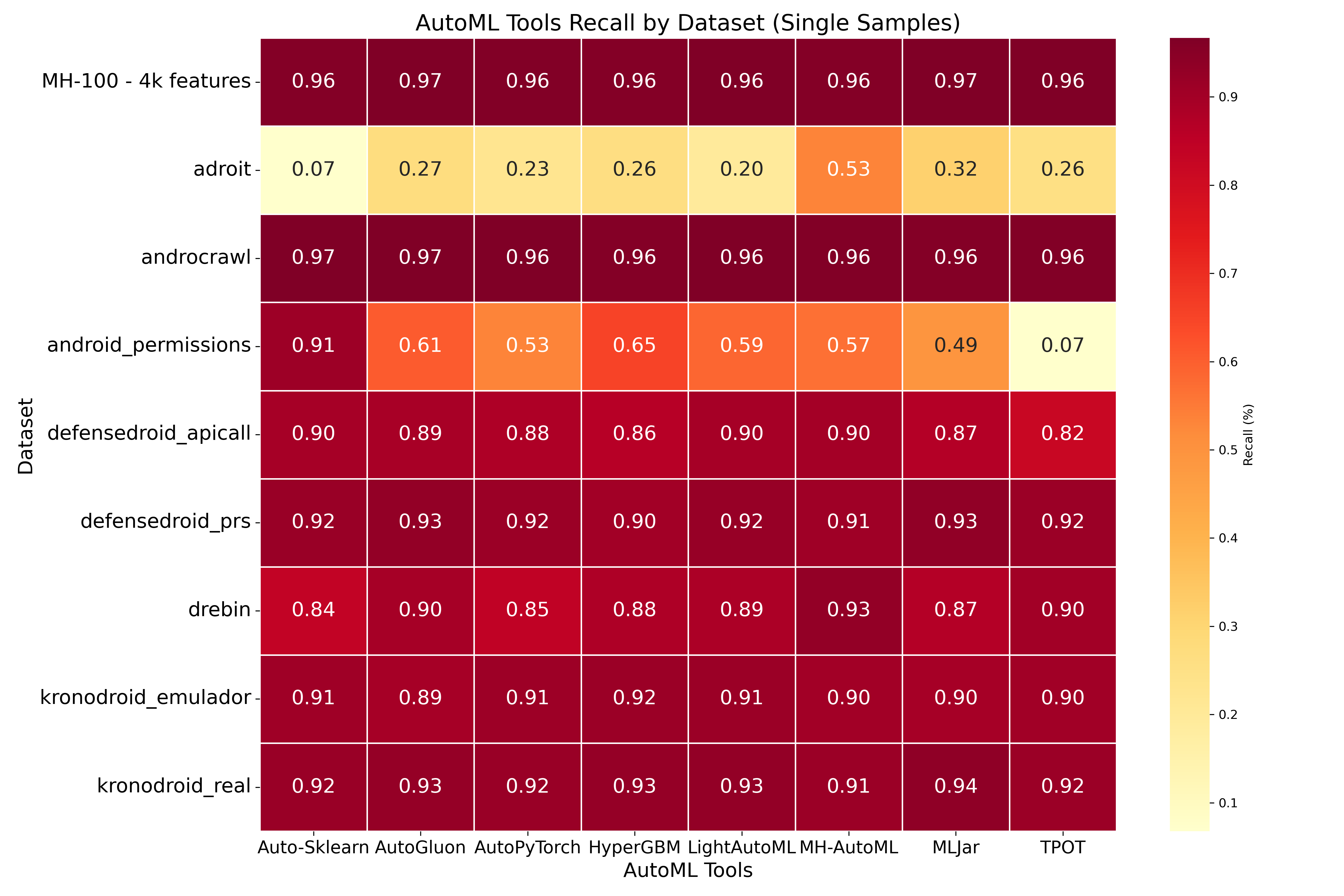}
\vspace{0.1cm}
\caption{Recall performance across balanced datasets for all evaluated AutoML tools.}
\label{fig_desempenho_recall_balanceados}
\end{figure*}

\begin{figure*}[!htb]
\centering
\includegraphics[width=0.75\textwidth]{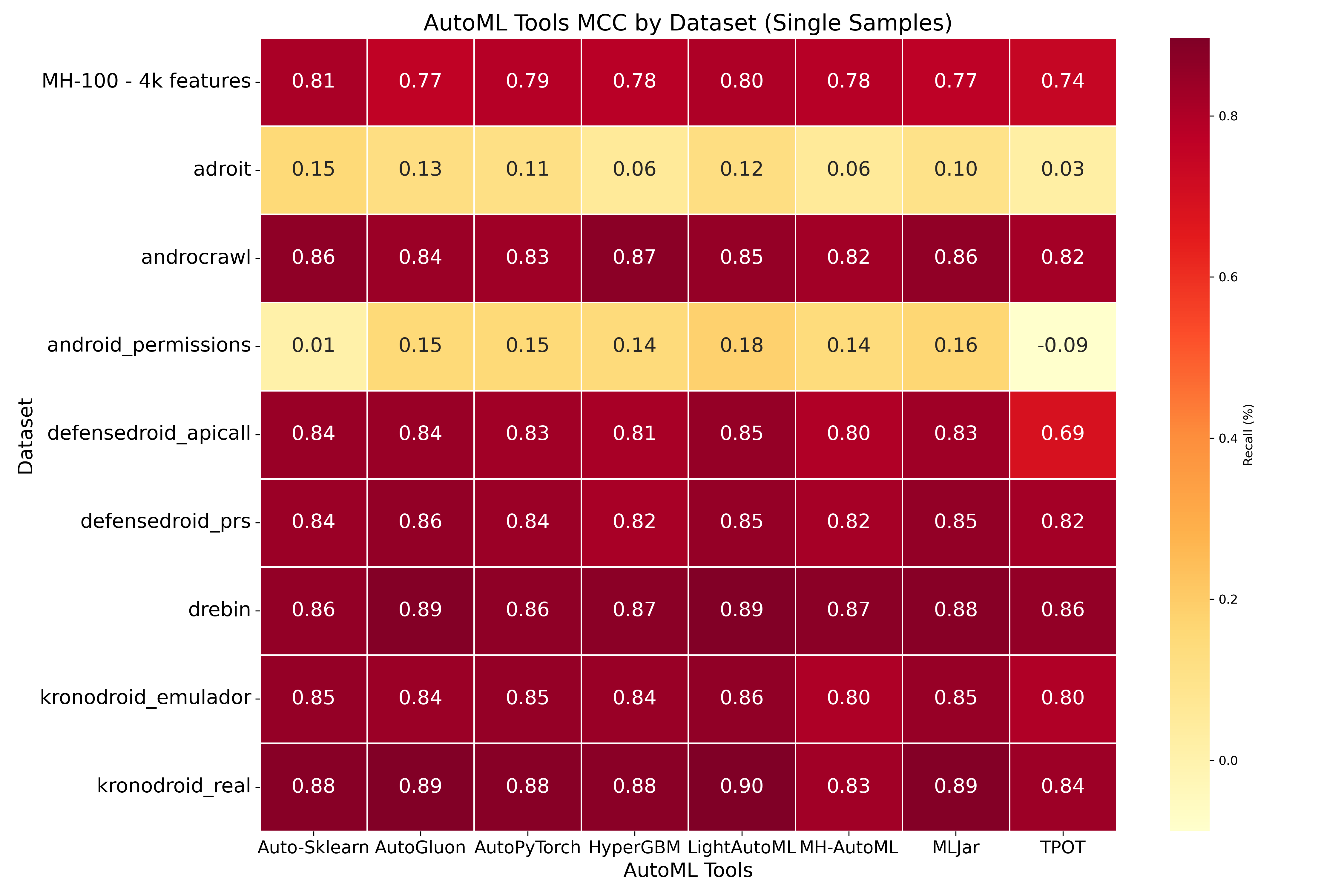}
\vspace{0.1cm}
\caption{MCC performance across balanced datasets for all evaluated AutoML tools.}
\label{fig_desempenho_mcc_balanceados}
\end{figure*}

\begin{figure*}[!htb]
\centering
\includegraphics[width=0.75\textwidth]{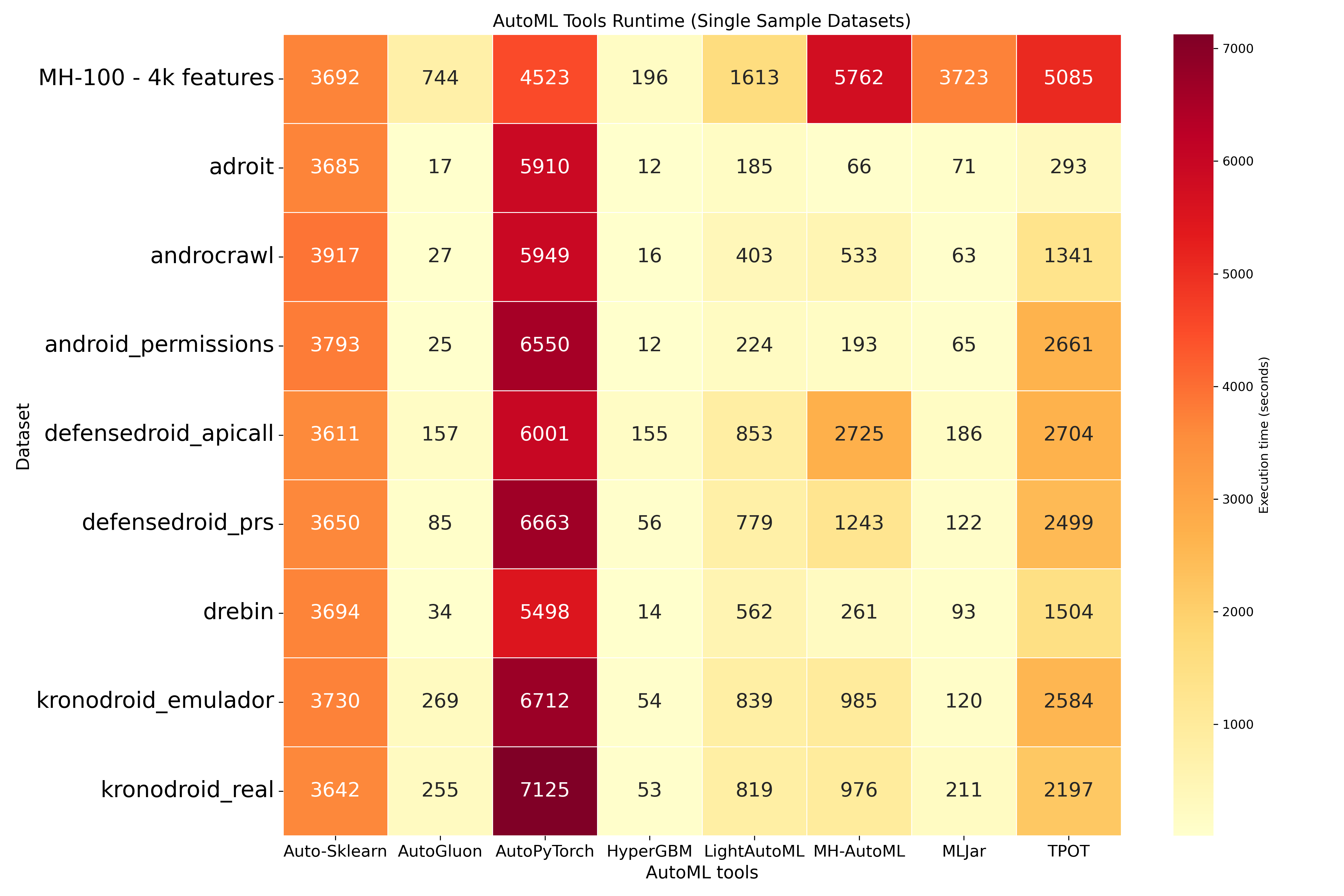}
\vspace{0.1cm}
\caption{Execution times across balanced datasets for all evaluated AutoML tools.}
\label{fig_tempo_exec_balanceados}
\end{figure*}

We evaluate the performance of 8 AutoML tools across 9 balanced datasets using Recall, MCC, and execution time metrics. Figures \ref{fig_desempenho_recall_balanceados}, \ref{fig_desempenho_mcc_balanceados}, and \ref{fig_tempo_exec_balanceados} present these results through heatmap visualizations that enable direct comparison between tools and datasets.

Our analysis reveals distinct performance patterns across different datasets. LightAutoML demonstrates consistent results, achieving 91.49\% recall and 86.09\% MCC in the Kronodroid Emulator dataset with an efficient 13 minute 59 second runtime. This strong performance continues in the Kronodroid Real dataset where it achieves 93.16\% recall and 89.66\% MCC, showing its reliability with balanced data.

The Androcrawl dataset shows uniformly high recall scores between 95\% and 97\% across all tools, with AutoGluon leading at 96.56\%. HyperGBM achieves the best MCC (87.20\%) in just 16 seconds, suggesting it can be effective for certain balanced datasets despite its generally weaker performance elsewhere.

Some datasets present particular challenges. The Android Permissions dataset shows poor performance across tools, with Auto-Sklearn's 91.41\% recall accompanied by a near-zero MCC (1.09\%), indicating significant false positive issues. TPOT performs especially poorly here with negative MCC (-8.80\%), suggesting fundamental classification problems.

For the DefenseDroid datasets, we observe MLJar and AutoGluon performing best in the PRS version (93\% recall, 85\% MCC), while MH-AutoML leads in recall (89.75\%) for the API Calls version with LightAutoML showing better MCC (85.26\%). The Drebin-215 dataset shows MH-AutoML leading in recall (93.10\%) and LightAutoML in MCC (89.18\%).

The Adroit dataset proves challenging for all tools, with MH-AutoML's 53.19\% recall being the highest but still indicating classification difficulties. In the MH-100 dataset, AutoGluon achieves highest recall (96.68\%) with reasonable runtime (12m24s), while LightAutoML provides best balance (96.38\% recall, 80.07\% MCC) in 26m53s.

The performance variations reflect the tools' underlying architectures. LightAutoML's blending technique and early stopping provide speed and accuracy. AutoGluon's weighted ensembles and fixed hyperparameters offer similar benefits. HyperGBM's limited model selection prioritizes speed, while Auto-Sklearn and TPOT incur time penalties from complex optimization without proportional accuracy gains.

Our evaluation reveals important trade-offs between performance metrics and computational efficiency across the AutoML tools. The results show that different tools excel in different aspects of malware detection, with no single solution dominating all categories.

MH-AutoML demonstrates strong recall capabilities (95.66\%) with reasonable MCC (78.36\%), though its execution time of 1 hour and 36 minutes makes it less suitable for time-sensitive applications. This performance pattern suggests the tool may be particularly effective when false negatives carry higher costs than false positives in malware detection scenarios.

Auto-Sklearn shows comparable recall (95.66\%) to MH-AutoML with slightly better MCC (80.95\%) and moderately faster execution (1 hour and 1 minute). The tool appears to strike a better balance between precision and recall, though its runtime remains substantial compared to faster alternatives.

TPOT presents an interesting case with the highest recall (96.43\%) but the lowest MCC (74.30\%) among the top performers, coupled with a long execution time (1 hour and 24 minutes). This discrepancy suggests the tool may be over-optimizing for sensitivity at the expense of overall classification quality.

AutoPyTorch delivers consistent performance across metrics (96.02\% recall, 78.81\% MCC) with execution times (1 hour and 15 minutes) similar to other intermediate-speed tools. Its balanced results indicate it may be a reliable choice when both sensitivity and specificity are important.

LightAutoML emerges as one of the top performers, achieving excellent recall (96.38\%) and the best MCC (80.07\%) among all tools while maintaining reasonable execution times (26 minutes and 53 seconds). This combination of accuracy and efficiency makes it particularly suitable for practical malware detection applications.

HyperGBM stands out for its exceptional speed (3 minutes and 16 seconds), though this comes at the cost of slightly lower performance metrics (95.66\% recall, 77.86\% MCC). The tool may be preferable when rapid analysis is more critical than maximum accuracy.

MLJar delivers competitive results (96.58\% recall, 76.88\% MCC) with intermediate execution times (1 hour and 2 minutes), positioning it as a viable alternative when both accuracy and runtime need to be considered.

Furthermore, our evaluation reveals several important patterns in tool performance that can guide selection for malware detection tasks. We observe that different tools excel in different aspects, with trade-offs between accuracy metrics and computational efficiency that must be carefully considered for each application scenario.

LightAutoML and AutoGluon emerge as particularly well-balanced options, providing strong performance across both accuracy metrics and execution time. LightAutoML achieves this through its efficient pipeline architecture that combines multiple base models (CatBoost, LightGBM, XGBoost, Random Forest, and Neural Networks) using a blending technique similar to stacking. The tool implements \textit{early stopping} and parallel optimization to maintain competitive results while reducing execution times. AutoGluon employs a different but equally effective approach, using weighted ensembles of base models with fixed hyperparameters to balance performance and speed.

For applications where recall is the primary concern, MH-AutoML and TPOT demonstrate superior sensitivity in detecting malware samples. However, these gains come with notable trade-offs - MH-AutoML requires significantly longer execution times (1 hour 36 minutes in our tests) due to its feature selection process, while TPOT shows lower MCC scores indicating potential issues with false positives. These tools may be most appropriate when the cost of missing actual malware outweighs the impact of occasional false alarms.

HyperGBM stands out for time-sensitive applications, completing analyses in just 3 minutes 16 seconds - the fastest among all evaluated tools. This speed comes from its focused approach of training only four core models (XGBoost, CatBoost, LightGBM, and HistGradientBoosting) with fixed parameters by default. While its accuracy metrics are slightly lower than the top performers, the dramatic time savings may justify this trade-off in scenarios requiring rapid analysis.

The tools showing longest execution times - AutoPyTorch (1h15m), Auto-Sklearn (1h1m), and TPOT (1h24m) - employ complex optimization strategies that do not consistently translate to superior performance. Auto-Sklearn uses meta-learning with Bayesian optimization, AutoPyTorch relies on Bayesian optimization, and TPOT implements genetic programming. While these approaches can theoretically discover optimal configurations, in practice they often require substantial computational resources without delivering proportionally better results compared to more efficient alternatives.

Our analysis of the performance across different datasets reveals that tool behavior can vary significantly depending on data characteristics. As shown in Figures \ref{fig_desempenho_recall} and \ref{fig_desempenho_mcc_balanceados}, certain datasets like Adroit and Android Permissions prove challenging for all tools, suggesting these may require specialized approaches beyond current AutoML capabilities. The execution time patterns in Figure \ref{fig_tempo_exec_balanceados} demonstrate that while some tools maintain consistent speed across datasets (e.g., AutoGluon and HyperGBM), others show more variability in their runtime performance.

These findings suggest that optimal tool selection depends on several factors:
\begin{itemize}
\item The relative importance of recall versus MCC in the specific application context;
\item Available computational resources and time constraints;
\item Characteristics of the target dataset and malware features;
\item Tolerance for potential false positives versus false negatives.
\end{itemize}

The comprehensive performance data we provide enables practitioners to make informed decisions based on their specific requirements and constraints, rather than relying on general claims about tool superiority. Each tool's architectural approach leads to different performance characteristics that may make it more or less suitable for particular malware detection scenarios.

\subsection{Comparative Analysis: Balanced vs Original Datasets}
\label{sec_comparative_analysis}

We examine the impact of data balancing on tool performance by comparing results between original and balanced datasets. Our analysis reveals several important patterns in how different AutoML tools respond to class balancing techniques.

The KronoDroid Emulator dataset shows improved recall after balancing, with HyperGBM increasing from 94.03\% to 91.69\%. However, this recall improvement comes with a slight decrease in MCC for most tools, suggesting that while balancing helps detect more positive cases, it may reduce overall classification precision. LightAutoML maintains robust performance in both metrics (91.49\% recall, 86.09\% MCC), demonstrating its stability across different data distributions.

In the Androcrawl dataset, we observe significant recall improvements after balancing, with AutoGluon increasing from 91.00\% to 96.56\%. MH-AutoML also performs well with 96.13\% recall. Interestingly, MCC values remain relatively stable in this dataset, indicating that the correlation between predictions and actual labels is less affected by balancing compared to other datasets.

The Android Permissions dataset presents a unique case where balancing produces mixed results. Auto-Sklearn achieves 91.41\% recall compared to 99.19\% in the original data, but with an extremely low MCC of 0.0109, suggesting severe issues with false positives. TPOT performs particularly poorly on the balanced version, with recall dropping to 6.79\% and negative MCC values, indicating fundamental compatibility issues with balanced data.

For the Drebin dataset, MH-AutoML shows strong recall (93.10\%) on balanced data compared to 98.51\% originally, while LightAutoML maintains the highest MCC (89.18\%) among all tools. This suggests that some tools can adapt well to balanced distributions while preserving classification quality.

Execution times generally decrease after balancing, with AutoGluon's KronoDroid Emulator processing time dropping from 17 minutes to 4 minutes 29 seconds. However, some tools like Auto-Sklearn and AutoPyTorch continue to require substantial computation time even with balanced data.

Our findings suggest that:
\begin{itemize}
\item Data balancing generally improves recall, especially for imbalanced datasets;
\item MCC may decrease slightly after balancing, indicating a precision trade-off;
\item Some tools (e.g., TPOT) perform poorly on balanced data;
\item Execution times typically decrease with balanced datasets;
\item LightAutoML and AutoGluon show consistent performance across both data versions.
\end{itemize}

These results indicate that while balancing can improve detection rates, practitioners should carefully consider:
\begin{itemize}
\item The specific requirements of their detection scenario;
\item The trade-offs between recall and precision;
\item The compatibility of their chosen tool with balanced data;
\item The computational resources available.
\end{itemize}

The choice between original and balanced approaches should be guided by both performance requirements and the characteristics of the specific tools being considered. Our comparative analysis provides concrete evidence to support these decisions across different dataset types and tool implementations.

\section{Conclusion}
\label{sec_conclusion}

Our comprehensive evaluation of MH-AutoML and seven other AutoML tools (AutoGluon, Auto-Sklearn, TPOT, AutoPyTorch, MLJAR, HyperGBM, and LightAutoML) for Android malware detection reveals several important findings about their respective strengths and trade-offs.

MH-AutoML distinguishes itself through advanced interpretability and transparency features that surpass other tools in our evaluation. The tool incorporates specialized feature selection methods and model-specific approaches for Android malware classification, along with integrated versioning and tracking capabilities through MLflow. These features enable detailed analysis of the complete AutoML pipeline, making MH-AutoML particularly valuable for research and debugging scenarios. While its current version shows promising results that compete with established tools, there remains potential for further improvements in future iterations.

When examining transparency and interpretability aspects (Figure \ref{fig:AutoML_Feature_Score}), we find that both MH-AutoML and MLJAR provide user-friendly interfaces suitable for non-expert users. MH-AutoML extends this advantage further with its comprehensive experiment tracking through MLflow, offering researchers and practitioners unprecedented visibility into the model development process. Although these tools may not achieve the absolute highest performance metrics in detection tasks, they deliver consistent and reliable results across all tested datasets while maintaining reasonable execution times. This combination makes them particularly suitable for users without extensive machine learning expertise.

Performance analysis reveals distinct patterns across different metrics. In terms of execution speed, HyperGBM and AutoGluon emerge as the fastest options. HyperGBM achieves this through various optimization techniques, including early stopping to prevent overfitting and reduce training time. AutoGluon takes a different approach by using fixed hyperparameters across its 14 models, eliminating the time-consuming optimization phase while still delivering competitive results.

For classification performance measured through accuracy metrics, AutoGluon and LightAutoML consistently produce the best results across the diverse datasets in our study. LightAutoML stands out as the most balanced tool overall, offering strong performance in both accuracy metrics and execution times. AutoGluon, HyperGBM, MH-AutoML, and MLJar also demonstrate good trade-offs between performance and efficiency, though each with different strengths. On the other hand, AutoPyTorch, Auto-Sklearn, and TPOT prove less competitive in our evaluation, particularly when considering time efficiency alongside classification performance.

These findings suggest that tool selection should be guided by specific use case requirements:
\begin{itemize}
\item For maximum interpretability and transparency: MH-AutoML or MLJAR;
\item For fastest execution: HyperGBM or AutoGluon;
\item For best classification performance: AutoGluon or LightAutoML;
\item For balanced overall performance: LightAutoML.
\end{itemize}

The results demonstrate that while no single tool dominates all categories, the current landscape of AutoML solutions offers viable options for various Android malware detection scenarios, each with different strengths and trade-offs that practitioners can leverage based on their specific needs.

\section*{Declarations}

\subsection*{Contributions}
Joner Assolin and Diego Kreutz contributed to the conception and design of this study. Joner Assolin and Gabriel Canto conducted the experiments. Diego Kreutz, Hendrio Bragança, Eduardo Feitosa, Angelo Nogueira and Vanderson Rocha provided critical revisions, supervision, and feedback. Angelo Nogueira was one of the primary authors of Section 3. Joner Assolin, Gabriel Canto, and Diego Kreutz were the main contributors and writers of the manuscript. All authors reviewed and approved the final version.

\subsection*{Interests}
The authors declare that they have no competing interests.

\subsection*{Funding}
This research was partially funded, as stipulated in Articles 21 and 22 of Decree No. 10.521/2020, under Federal Law No. 8.387/1991, through agreement No. 003/2021, signed between ICOMP/UFAM, Flextronics da Amazônia Ltda., and Motorola Mobility Comércio de Produtos Eletrônicos Ltda. This work was also supported by the Coordination for the Improvement of Higher Education Personnel – Brazil (CAPES) – Financing Code 001, and partially supported by the Amazonas State Research Support Foundation (FAPEAM) through the POSGRAD project 2024/2025. This research was also partially funded by FAPERGS through grant agreements 24/2551-0001368-7 and 24/2551-0000726-1.

\subsection*{Materials}
The source code and datasets analyzed in this study are publicly available in the MH-FSF GitHub repository: \url{https://github.com/SBSegSF24/MH-FSF}.

\bibliographystyle{IEEEtran}
\bibliography{refs}

\end{document}